\documentclass[twocolumn]{revtex4-1}
\usepackage{natbib}
\usepackage{amsmath}
\usepackage{graphicx}
\begin{document}
\title{Ideal near-field thermophotovoltaic cells}
\author{Sean Molesky}
\author{Zubin Jacob}
\email{zjacob@ualberta.ca} 
\affiliation{University of Alberta, Department of Electrical and Computer Engineering \\ T6G 2V4 Edmonton, Alberta, Canada}
%%%%%%%%ABSTRACT
\begin{abstract}
We ask the question, what are the ideal characteristics of a near-field thermophotovoltaic cell? Our search leads us to a reformulation of near-field radiative heat transfer in terms of the joint density of electronic states of the emitter-absorber pair in the thermophotovoltaic system. This form reveals that semiconducting materials with narrowband absorption spectra are critical to the energy conversion efficiency. This essential feature is unavailable in conventional bulk semiconductor cells but can be obtained using low dimensional materials. Our results show that the presence of matched van Hove singularities resulting from quantum-confinement in the emitter and absorber of a thermophotovoltaic cell boosts both the magnitude and spectral selectivity of radiative heat transfer; dramatically improving energy conversion efficiency. We provide a model near-field thermophotovoltaic system design making use of this idea by employing the van Hove singularities present in carbon nanotubes. Shockley-Queisser analysis shows that the predicted heat transfer characteristics of this model device are fundamentally better than existing thermophotovoltaic designs. Our work paves the way for the use of quantum dots, quantum wells, two-dimensional semiconductors, semiconductor nanowires and carbon nanotubes as future materials for thermophotovoltaic cells.
\end{abstract}

\maketitle
\section{Introduction}
The thermophotovoltaic (TPV) method generalizes the concept of conventional photovoltaics by replacing the source solar spectrum with the thermal emission of an engineered selective emitter\cite{Wedlockt1963}. When heated, the emitter is designed to deliver photonic power over a narrow spectral window which can be efficiently converted into electrical power by a matched photovoltaic cell, assumed to be near room temperature. In this way, electrical power can be extracted from any heat reservoir\cite{Nelson2003,Basu2007}. In principle, the TPV approach avoids the two primary loss mechanisms of single-junction solar photovoltaics: the generation of sub-bandgap photons, and the thermalization of excitons with energy much greater than the bandgap\cite{Shockley1961}. For this reason, the theoretical power conversion limitations of TPVs are near that of an ideal heat engine\cite{Harder2003,Datas2010a}.\par
However, in practice, implementation of the TPV idea is difficult. The spectral irradiance of any emitter in the far-field is bounded by that of a blackbody with equal temperature\cite{Greffet1998}, see Fig.1. For an emitter with temperature below approximately 1000 K, the fraction of emitted power in the range of contemporary low bandgap photovoltaics $(0.4-0.75\,$eV$)$ is negligible\cite{Bermel2010}, making practical photovoltaic conversion problematic. To overcome this limitation, the emitter of current TPV systems must operate at a temperature between roughly 1200 and 1500 K, creating severe design requirements. Specifically, the emitter must be thermally robust in this temperature range, suppress emission below the bandgap of the photovoltaic cell, which constitutes the majority of the solar spectrum, and have emissivity near the blackbody limit above the bandgap\cite{Rephaeli2009,Celanovic2008}. Even without considering temperature induced degradation of optical properties and structure, current theoretical proposals still require emitter temperatures surpassing 1500 K to reach 40\% conversion efficiency under Shockley-Queisser analysis\cite{nagpal2008efficient,Molesky2012,Nam2014}.\par
\begin{figure}[h!]
 \vspace{-6pt}
\centering
\includegraphics{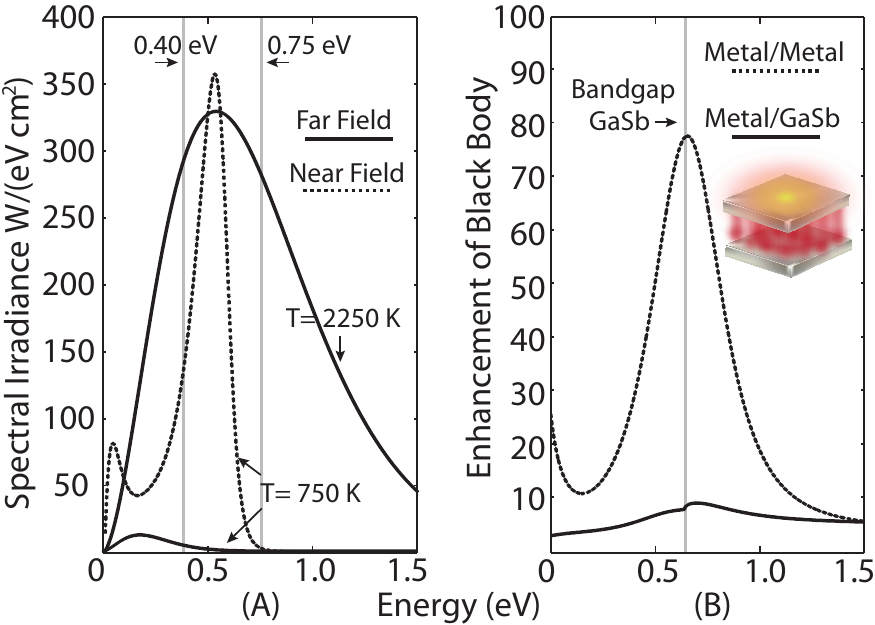}
 \vspace{-15pt}
\caption{(A) Comparison of radiative heat transfer between the near-field and far-field calculated using \eqref{RytovVanHovePolder}. The near-field system consists of two half-spaces with optical response of a hypothetical Drude metal, one at 750 K, the other at 0 K, with 0.08 eV loss parameter, 0.58 eV plasma frequency, and an additional static imaginary permittivity of 1, separated by a 100 nm gap. This arrangement is shown schematically as an inset in panel (B). The vertical gray lines mark an approximate range of current single-junction low bandgap photovoltaics. The enhancement of radiative heat transfer through evanescent modes relaxes the necessity of maintaining the emitter at extremely high temperature. (B) Comparison of the enhancement of radiative heat transfer beyond the blackbody limit for half-spaces separated by a 100 nm gap, shown schematically as an inset. The enhancement is temperature independent. The dashed curve is found by assuming identical hypothetical Drude metals with 0.21 eV loss parameter, 0.83 eV plasma frequency, and an additional static imaginary permittivity of 2.5. The solid curve results when one of the metal half-spaces is replaced by a gallium antimonide photovoltaic cell. The inclusion of the bulk photovoltaic cell  greatly reduces the enhancement effects provided by operating in the near-field.}
 \vspace{0 pt}
\end{figure}
An intriguing prospect to partially alleviate this high emitter temperature requirement is to bring the matched emitter and absorber into the near-field; where the presence of tunneling photons allows heat transfer to surpass the far-field blackbody limit (near-field thermophotovoltaics, NFTPV\cite{DiMatteo2001,Narayanaswamy2003,Laroche2006}). Making use of matched near-field resonances\cite{Ben-Abdallah2009,Biehs2013}, for example surface plasmon polaritons\cite{Ilic2012,boriskina2013plasmonic} as in Fig.1, radiative heat transfer can be made spectrally sharp and significantly greater than the far-field limit (Super-Planckian emission\cite{Guo2012}).\par
Nevertheless, basic design issues persist even in theoretical consideration. Analysis of near-field radiative heat transfer has conventionally focused on the maximization of total transfered power, leading naturally to designs using coupled modes of symmetric structures. However, to extract usable electrical power, one half of the near-field resonant architecture must function as a photovoltaic cell. And when realistic bulk semiconductors are included as the low temperature absorber, radiative heat transfer is greatly reduced and spectrally broadened\cite{Messina2013,bright_performance_2014}. For example, considering a 100 nm gap, and identical hypothetical Drude metals with surface plasmon polariton resonances matched to the bandgap of gallium antimonide photovoltaic cell\cite{Adachi1989}, replacing the symmetric metal absorber with the photovoltaic cell reduces radiative heat transfer by a factor of nearly 10, Fig.1. Furthermore, although the temperature difference between the emitter and the photovoltaic cell can be comparatively reduced, bringing the photovoltaic cell into the near-field of the emitter will, in practice, increases its operating temperature and thus reduces its conversion efficiency\cite{Francoeur2011}. Resultantly, much of the performance gain offered by moving to the near field thermophotovoltaic systems is lost.\par
\begin{figure}[ht!]
 \vspace{0pt}
\centering
\includegraphics{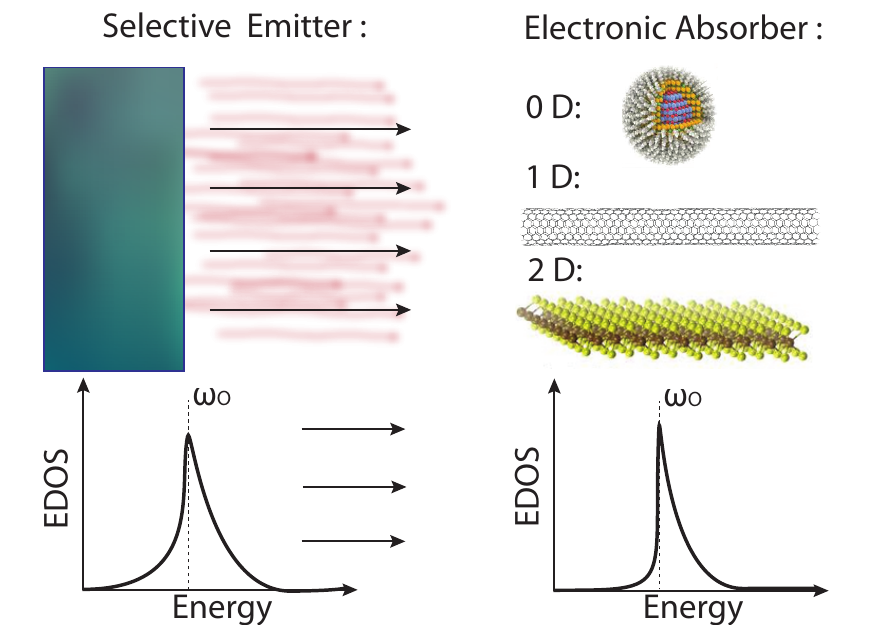}
 \vspace{-10pt}
\caption{Schematic illustration of the central result of this article. Here, EDOS stands for the joint density of electronic states. We will show that by matching peaks in the joint density of electronic states of the emitter to peaks in the joint density of electronic states of the photovoltaic cell the spectral selectivity of near-field radiative heat transfer and efficiency of near-field thermophotovoltaics can be greatly improved.}
 \vspace{0pt}
\end{figure}
In this article, we show that the narrowband optical absorption arising from quantum confinement leads to spectrally selective and efficient near-field heat transfer, see Fig.2. This is a fundamental advantage since absorbed thermal power is the source of exciton generation. Simultaneously, we also uncover the root cause of the poor performance of traditional bulk semiconductor photovoltaic cells in the near-field by recasting near-field radiative heat transfer in terms of the joint density of electronic states. Appealingly, this results requires only simple arguments from Shockley-Queisser analysis and the essential Kramers-Kronig optical response constraints. We also provide the basis for an experimentally realizable near-field thermophotovaltic system making use of carbon nanotubes, and calculate its enhanced performance metrics. Our predictions can be verified by near-field thermal emission spectroscopy, providing a road map for the use of quantum dots, quantum wells, two-dimensional semiconductors, semiconductor nanowires and carbon nanotubes as the future building blocks of thermophotovoltaic devices. 
\section{Energy considerations for maximal radiative heat transfer}
From the pioneering works of Rytov\cite{Rytov1984}, and Polder and van Hove\cite{Polder1971}, the spectral irradiance between two planar half spaces separated by a vacuum gap is given by
\begin{equation}
\begin{split}
\left<Q\left(T_{1},T_{2},\omega_{o}\right)\right>=\frac{\omega_{o}^2\left|\Theta\left(T_{1},\omega_{o}\right)\right)-\Theta\left(T_{2},\omega_{o}\right)|}{4\pi^{2} c^2}\\
\Bigg(\sum\limits_{j=s,p}\int\limits_{v}^{1}dk_{\rho}\frac{\left(1-\left|r_{1}^{\,j}\right|^2\right)\left(1-\left|r_{2}^{\,j}\right|^2\right)}{\left|1-r_{1}^{\,j}r_{2}^{\,j}e^{-2i\,k_{\perp}^{v}d}\right|^2}k_{\rho}+\\
4\int\limits_{1}^{\infty}dk_{\rho}\frac{Im\left[r_{1}^{\,j}\right]Im\left[r_{2}^{\,j}\right]e^{-2Im\left[k_{\perp}^{v}d\right]}}{\left|1-r_{1}^{\,j}r_{2}^{\,j}e^{-2i\,k_{\perp}^{v}d}\right|^2}k_{\rho}\Bigg).
\label{RytovVanHovePolder}
\end{split}
\end{equation}
Where $c$ is the speed of light in vacuum, $\Theta\left(T_{i},\omega_{o}\right)$ the mean energy expectation value of the canonical harmonic oscillator at temperature $T_{i}$ and angular frequency $\omega_{o}$, $d$ the vacuum gap separation normalized by the magnitude of the free space wave vector at $\omega_{o}$, $T_{i}$ the temperature of the $i$th half space, $k_{\rho}$ the  parallel component of the wave vector, again normalized by the magnitude of the free space wave vector at $\omega_{o}$, $r^{\,j}_{i}$ the $j$-polarized reflection coefficient of the $i$th half space, $k_{\perp}^{v}$ the normalized perpendicular component of the wave vector in the vacuum gap, and Im$\left[...\right]$ the imaginary part of the enclosed function.\par
Intuitively, radiative heat transfer between planar half spaces is maximized when the structures are symmetric, and the vacuum gap is made vanishingly small, $d\rightarrow0$. In this limit the integrand of the spectral irradiance, $H\left(T_1,T_2,\omega_{o},k_{\rho}\right)$ from Eq.\eqref{RytovVanHovePolder}, is dominated by the high momentum p-polarized evanescent contribution, $k_{\rho}\gg 1$, and is well approximated by the bound form
\begin{equation}
H\left(T_1,T_2,\omega_{o},k_{\rho}\right)=\frac{\omega_{o}^2\left|\Theta\left(T_{1},\omega_{o}\right)\right)-\Theta\left(T_{2},\omega_{o}\right)|}{4\pi^{2} c^2\left(1+\left(\epsilon^{'}\!\!\left(\omega_{o}\right)/\epsilon^{''}\!\!\left(\omega_{o}\right)\right)^{2}\right)}k_{\rho};
\label{equation_1}
\end{equation}
 with $\epsilon\left(\omega_{o}\right)$ denoting the relative permittivity of both media, and the $'$ and $''$ superscripts marking the real and imaginary parts of the function. To achieve the greatest possible radiative heat transfer between identical media at given wave vector and frequency, the so called upper bound of radiative heat transfer per channel\cite{Pendry1999}, the ratio 
\begin{equation}
\gamma\left(\omega_{o}\right)\!:=\frac{\epsilon^{'}\!\!\left(\omega_{o}\right)}{\epsilon^{''}\!\!\left(\omega_{o}\right)},	
\label{equation_2}
\end{equation}
must be made as small as possible. (This condition is antithetical to the plasmonic field enhancement figure of merit where $\gamma\left(\omega_{o}\right)$ is maximized.) \par      
Applying the Kramers-Kronig relation\cite{landau1984electrodynamics} to the real part of the permittivity, the $\gamma\left(\omega_{o}\right)$ factor can be expressed entirely in terms of $\epsilon^{''}\!\!\left(\omega_{o}\right)$ as
\begin{equation}
\gamma\left(\omega_{o}\right)=\frac{1}{\epsilon^{''}\!\!\left(\omega_{o}\right)}+\frac{2}{\pi\epsilon^{''}\!\!\left(\omega_{o}\right)}\;{\mathcal P}\int\limits_{0}^{\infty} d\omega\;\frac{\omega\;\epsilon^{''}\!\!\left(\omega\right)}{\omega^{2}-\omega^{2}_{o}}.
\label{equation_3}
\end{equation}
For energy harvesting with semiconductor photovoltaic cells, Shockley-Queisser efficiency analysis can be used to provide constraints to the material parameters which minimize $\gamma\left(\omega_{o}\right)$: (1) No radiative heat transfer should occur below the bandgap frequency $\omega_{g}$ because it cannot be converted into useful electrical power.  Mathematically this criterion is stated as $\gamma\left(\omega_{o}\right)\!\rightarrow\!\infty$ for all $\omega_{o}<\omega_{g}$, equivalent to $\epsilon^{''}\!\!\left(\omega_{o}\right)\rightarrow 0$ for all $\omega_{o}<\omega_{g}$. The dispersive part of $\gamma\left(\omega_{g}\right)$ then provides an explicitly positive contribution, and its minimization requires $\epsilon^{''}\!\!\left(\omega_{o}\right)\rightarrow 0$ for all $\omega_{o}\neq\omega_{g}$. This restriction on the global absorption characteristics of the media, $\epsilon^{''}\!\!\left(\omega_{o}\right)$, can be immediately translated into two statements concerning its polarization, $\epsilon^{'}\!\!\left(\omega_{o}\right)$. First, $\epsilon^{'}\!\!\left(\omega_{g}\right)>0$. Second, the polarization of the media acts as a store for photonic power. From this store the initially transferred power may either be converted into the internal degrees of freedom of the absorber or return to the emitter, limiting the total transferred power. The larger the polarization of the medium is the smaller radiative heat transfer will be.\par
(2) radiative heat transfer should be made as great as possible at the bandgap frequency, $\epsilon^{''}\!\!\left(\omega_{g}\right)\rightarrow\infty$. This second requirement for achieving $\gamma\left(\omega_{g}\right)\rightarrow 0$ is provided by the fluctuation dissipation theorem. To maximize radiative heat transfer at $\omega_{g}$, we require $\epsilon^{''}\!\!\left(\omega_{g}\right)\rightarrow\infty$ to generate the largest possible thermal currents, the first term of Eq.\eqref{equation_3}. The balance of these two conditions leads unequivocally to the conclusion that to minimize $\gamma\left(\omega_{g}\right)$, under the above Shockley-Queisser constraints, $\epsilon^{''}\!\!\left(\omega_{o}\right)$ must be sharply resonant about $\omega_{g}$. Strictly, simultaneous minimization of $\gamma\left(\omega_{g}\right)$ and complete suppression of heat transfer for all sub-bandgap frequencies is only possible if $\epsilon^{''}\!\!\left(\omega_{o}\right)$ is mathematically equivalent to the Dirac delta distribution, $\epsilon^{''}\!\!\left(\omega_{o}\right)\rightarrow\delta\left(\omega_{g}\right)$. \par
The implications of this result to near-field energy harvesting become immediate by recalling the intimate connection between optical dissipation, and the joint density of electronic states,
\begin{equation}
\epsilon^{''}\!\!\left(\omega_{o}\right)=\frac{e^2}{\pi m_{e}^{2}\omega_{o}^{2}}\int dS^{d}\frac{\left|{\bf a}_{o}\cdot{\bf p}_{ji}\right|^{2}}{\left|{\bf \nabla}_{\bf k}\;{\mathcal E}_{ji}\left({\bf k}\right)\right|};
\label{equation_4}
\end{equation}
with ${\bf a}_{o}$ denoting the polarization vector of the electromagnetic excitation, ${\bf p}_{ji}$ the matrix element of the momentum operator, $\omega_{o}$ the angular frequency, ${\mathcal E}_{ji}\left({\bf k}\right)$ the constant energy surface between filled and excited states such that ${\mathcal E}_{ji}\left({\bf k}\right)=\hbar\omega_{o}$, $dS^{d}$ the infinitesimal constant energy surface element in $d$ dimensions, and $e$ and $m_{e}$ the mass and charge of an electron. Applying Eq.\eqref{equation_4} to model free particle semiconductor systems in one, two and three spatial dimension we recover the well known energy scaling behaviors $\epsilon^{''}\!\!\left(\omega_{o}\right)\!\propto\!\sqrt{\hbar\omega_{o}-E_{g}}\,(d\!\!=\!\!3),\,C\,(d\!\!=\!\!2),\,$and $ 1/\sqrt{\hbar\omega_{o}-E_{g}}(d\!\!=\!\!1)$, where $C$ is a constant, and $E_{g}$ the energy of the bandgap, matching the behavior of the joint density of electronic states. It is important to note that suppression/enhancement of electronic states at a given frequency is the key to decreasing/increasing optical absorption. However, the Kramers-Kronig relation introduces dependence on the global frequency characteristics through the polarization of the medium. Both these factors have to be taken into account in order to minimize the factor governing spectrally selective near-field heat transfer, $\gamma(\omega_{o})$.\par
As previously mentioned, the principal method for overcoming the Shockley-Queisser limit in TPV devices is to spectrally tailor the photonic output of the emitter to frequencies just above the bandgap of the photovoltaic cell. But, in the near-field this frequency range also corresponds to the poorest absorption characteristics for a typical bulk semiconductor. Near the bandgap the $\sqrt{\hbar\omega_{o}-E_{g}}$ scaling of the joint density of electronic states gives rise to both low absorption, $\epsilon^{''}\left(\omega_{g}\right)$, and large polarization, which can be seen by the positive contribution of principal value part in the Kramers-Kronig relationship $\left(\epsilon^{'}(\omega_0)=1+\frac{2}{\pi}\;{\mathcal P}\int_{0}^{\infty} d\omega\;\frac{\omega\;\epsilon^{''}\!\!\left(\omega\right)}{\omega^{2}-\omega^{2}_{o}}\right)$. These two factors lead to large $\gamma\left(\omega_{o}\right)$ factors detrimental to spectrally selective heat transfer when $\omega_{o}$ is near $\omega_{g}$, Fig.1. In short, the usual properties of a three dimensional semiconductor are in complete disagreement with those required for maximal radiative heat transfer as given by Eqn.\eqref{equation_1}.\par
In stark contrast, significant suppression of the $\gamma(\omega_{o})$ factor can be obtained if the absorption, $\epsilon^{''}\left(\omega_{o}\right)$, of the semiconductor cell is strongly peaked at the operating frequency. According to Eq.\eqref{equation_4}, such optical behavior is achieved when the joint density of electronic states becomes similar to the ideal Dirac distribution at the bandgap. Suppression of the electronic states below the operating frequency is necessary to decrease transfer of inefficient sub-bandgap photons, as dictated by Shockley-Queisser analysis. The suppression of states above the bandgap is necessary to decrease the polarization of the medium at the bandgap, as required by the Kramers-Kronig relations. These features can essentially be obtained through van Hove singularities, which occur in a free particle model in low spatial dimensions. This evolution is shown in Fig.3, where the $\gamma\left(\omega_{o}\right)$ factors for free particle semiconductor systems in one, two and three dimensions have been calculated. This structure in the density of electronic states and narrowband absorption behavior is characteristic of semiconductors with quantum-confined spatial dimensions such as quantum dots, nanotubes, and two-dimensional materials. We conclude based on the above analysis that systems possessing these properties are  ideally suited for achieving spectrally selective radiative heat transfer and can function as ideal near-field thermophotovoltaic cells. 
\begin{figure}[ht!]
 \vspace{-5pt}
\centering
\includegraphics{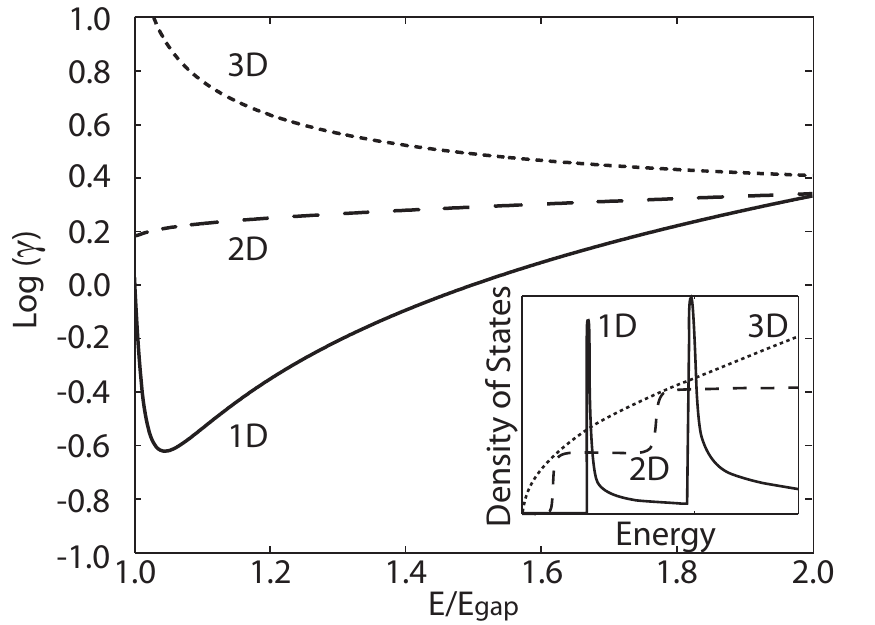}
 \vspace{-20pt}
\caption{: Base ten logarithm of the longitudinal $\gamma\left(\omega_{o}\right)\!:=\epsilon^{''}\left(\omega_{o}\right)/\epsilon^{''}\left(\omega_{o}\right)$ factor for one, two, and three dimensional model free particle semiconductors. The inset shows a schematic illustration of the evolution of the joint density of electronic states for zero, one, and two quantum-confined spatial dimensions characteristic of a semiconductor based on the work of K\"{u}mmel et al\cite{Lee1992}. By introducing dimensional constraints the dispersion of the joint density of electronic states can be altered to produce spectrally selectivity super-Planckian radiative heat transfer near the bandgap of a photovoltaic cell. This result illustrates both the primary drawback of employing traditional bulk semiconductor photovoltaics for near-field photonic energy conversion, and the usefulness of switching to photovoltaics with quantum-confined dimensions for this application.}
 \vspace{-10pt}
\end{figure}
\section{Model thermophotovoltaic system}
%The analysis of the previous section reveals that the magnitude and selectivity of near-field photonic power transfer for energy conversion applications can be greatly enhanced by employing photovoltaics with spectrally narrow absorption resonances. This type resonant optical behavior is one of the hallmarks of quantum confinement; making photovoltaic cells utilizing low dimensional materials such as quantum dots, quantum wells, graphene, two-dimensional semiconducotrs, semiconductor nanowires and carbon nanotubes\cite{hochbaum2009semiconductor,Fang2013,Shea2013} excellent candidates for future near-field TPV devices. As a framework of such a device, we consider here an idealized NFTPV system consisting of a emitter absorber pair utilizing carbon nanotubes\cite{jorio2007carbon} (CNTs) and calculate the radiative heat transfer characteristics of this model which could be observed in experiment.\par 

A wide body of previous work exists in the areas of photovoltaic cells that utilize materials such as quantum dots, graphene and carbon nanotubes \cite{hochbaum2009semiconductor,Fang2013,Shea2013}. Our analysis  in the previous section shows that any of these low dimensional material systems are excellent candidates for future near-field TPV cells. Again we emphasize that the efficiency boost arises from spectrally selective nature of the absorption in these systems which is the hallmark of quantum confinement.\par

As a framework of such a device, we consider here an idealized NFTPV system consisting of an emitter-absorber pair utilizing carbon nanotubes\cite{jorio2007carbon} (CNTs) and calculate the radiative heat transfer characteristics which can be observed in experiment. Our CNT model device is analogous to the one-dimensional free particle semiconductor example shown in Fig.3. The confinement of electrons in CNTs leads to strong van Hove singularities\cite{Cortes2013}, spectrally selective absorption, and semiconductor behavior for specific chiral vectors. Note that nearly identical performance would occur if the CNTs were exchanged for semiconductor nanowires\cite{Arnardottir2013}. We have focused on this CNT design, despite current challenges in fabrication and electron-hole pair collection\cite{wang2013carbon}, primarily because the thermal robustness\cite{ilic2012overcoming,Messina2013} and low bandgap of CNTs seems ideally suited to TPV applications.\par

Two versions of the CNT system have been considered. The first is designed for a 1300 K emitter and 300 K absorber; the second, for an 800 K emitter and 300 K absorber. In both cases, the absorber is composed of CNTs with chiral vector (19,0) embedded vertically in a matrix of zirconium dioxide on a tungsten backing. This choice of chiral vector allows the CNTs to operate as semiconductors, with a bandgap of 0.58 eV. This behavior occurs as the Dirac point of the band-structure is avoided\cite{wong2010carbon} due to the angular quantization. The exact current voltage characteristics of the CNT photovoltaic cells have not been included, and instead ideal p-n junction behavior has been assumed. This is in accordance with the Shockley-Quiesser analysis for ideal performance limitations. (To create fully functioning NFTPV cells the CNT on tungsten absorbers would be attached to external electronics and act as the photovoltaic elements\cite{bindl2013free,kymakis2002single,freitag2003}.)\par

Both emitters are composed of free standing CNTs with the same chiral vector as the CNTs considered for the absorber, and are set on a zirconium dioxide backing. The absorber and emitter are separated from each other by a 100 nm vacuum gap. In the high temperature case the fill fractions and thickness of the emitter and absorber CNT layers are 65\% - 50 nm, and 65\% - 20 nm. For the low temperature case these parameters are altered to 30\% - 50 nm, and 65\% - 100nm in the same order. A schematic of this setup is included as an inset in Fig.5.\par

The effective perpendicular permittivities of the two considered designs are shown in Fig.4. The absorption spikes, $\epsilon^{''}\left(\omega_{o}\right)$, can be tuned by varying the chiral vector. The first principles calculation of these optical properties is based on the Kubo formalism described by Falkovsky and Varlamov\cite{Falkovsky2007}, with additional loss included via the relaxation time approximation with an estimate relaxation time of one picosecond\cite{Kohler2011,Kohler2013}. The effective parallel permittivities, which play a secondary role in determining the transferred power, are calculated in identical manner. The main steps of this calculation are outlined in the appendix. Again, a variety of near-field emitters using other resonances, such as surface plasmon polaritons or phonon polaritons, or absorbers, such as those mentioned above, could also be used with the CNT absorber to obtain very similar results.\par
\begin{figure}[ht!]
 \vspace{-5pt}
\centering
\includegraphics{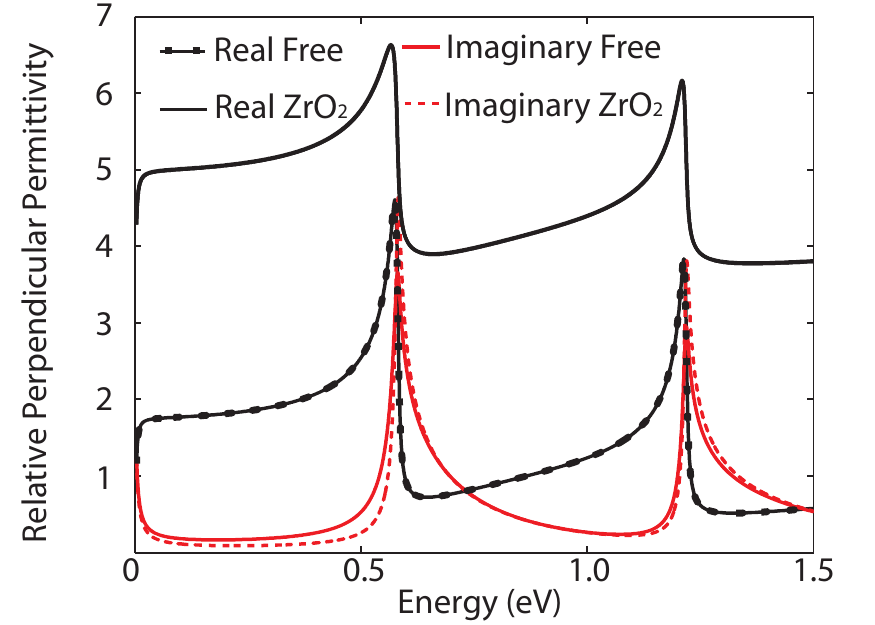}
 \vspace{-20pt}
\caption{: Effective relative permittivity perpendicular to the half spaces for the CNT emitter absorber pair designed to operate at 1300 K, depicted in Fig.5. For vacuum gaps on the order of hundreds of nanometers radiative heat transfer is dominated by tunneling photons with p-polarization. In this case, the overall radiative heat transfer properties are primarily determined by the perpendicular permittivity. As predicted by Eqn.(2), the radiative heat transfer
is maximized when the $\gamma\left(\omega_{o}\right)$ factor is minimized for both the emitter and absorber.}
 \vspace{-10pt}
\end{figure}
In figure 5, we plot the spectral irradiance for this CNT based emitter-absorber system calculated using Rytov's theory. As can be seen by comparing the perpendicular permittivity, Fig.4, with the spectral irradiance, Fig.5, the spectral position of peak radiative heat transfer corresponds strongly with the van Hove singularities and the minimization of $\gamma\left(\omega_{o}\right)$. In fact, although the analytical theory of \eqref{equation_2} is valid only in the very near-field, the spectral position of peak photonic transfer varies only 4\% as the vacuum gap is increased up to roughly half the wavelength of operation, 1 $\mu$m.\par
\begin{figure}[ht!]
 \vspace{-5pt}
\centering
\includegraphics{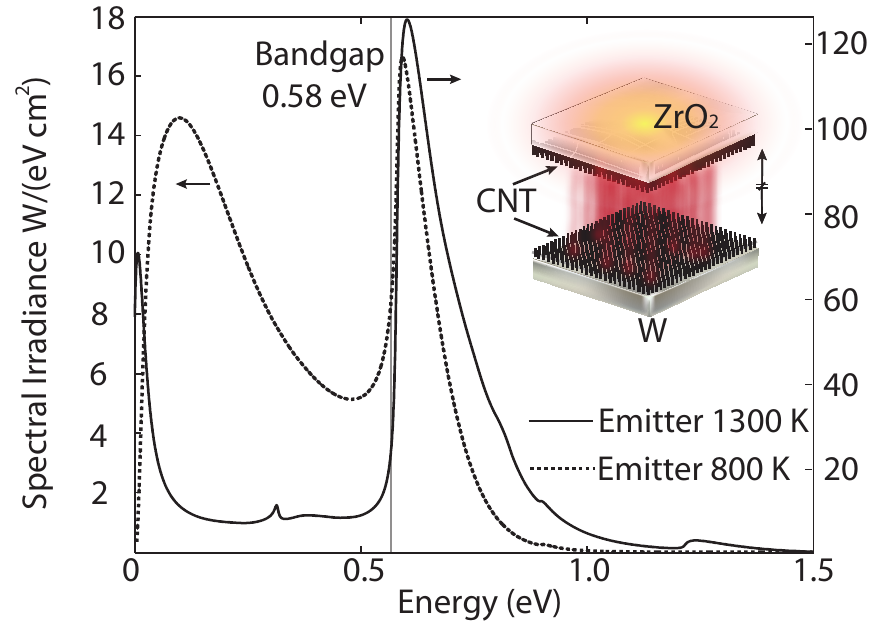}
 \vspace{-20pt}
\caption{: radiative heat transfer for two metamaterial emitter absorber pairs utilizing semiconducting CNTs optimized for emitter temperatures of 1300 K and 800 K. In both instances the absorber is assumed to be held at a temperature of 300 K. Parameters for the two systems are given in the main text. A schematic of the system is inset over the plot. The greatly enhanced, spectrally thin, radiative heat transfer peak produced slightly above the energy bandgap of the matched semi-conducting CNT absorber seen here is ideal for near-field photonic energy conversion applications.}
 \vspace{-10pt}
\end{figure}

The spectrally tailored nature of emission and absorption can lead to significant enhancement for energy conversion. Applying Shockley-Queisser analysis\cite{Shockley1961} as an upper-bound performance estimate, the CNT system with the 1300 K emitter (Fig. 5) could supply up to 15.00 W/cm$^{2}$ of electrical power with 53.0\% power conversion efficiency. Likewise, the 800 K emitter system could supply up to 1.07 W/cm$^{2}$ of electrical power with 15.6\% power conversion efficiency. For comparison, an optimized conventional bulk gallium antimonide\cite{Adachi1989} NFTPV design has also been considered. This system consists of a 175 nm thick Drude metal, with 0.74 eV plasma frequency, 0.21 eV loss parameter, on a tungsten backing with a vacuum gap separation of 100 nm. Under identical Shockley-Queisser analysis at a 1300 K emitter temperature this conventional system could produce 1.62 W/cm$^{2}$ of electrical power with 18.8\% power conversion efficiency, whereas with an emitter temperature of 800 K it could create only 0.01 W/cm$^{2}$ of electrical power with 1.6\% power conversion efficiency. More concretely, 73.1\% of photonic thermal power transferred from the emitter in the CNT NFTPV device lies above the bandgap, 20.7 W/cm$^{2}$, for the 1300 K model system and 25.0\%, 1.7 W/cm$^{2}$, in the 800 K model. In comparison, the optimized plasmonic system described above transfers 28.1\% of its thermal power above the bandgap of the gallium antimonide photovoltaic cell, 2.4 W/cm$^{2}$, when the emitter is given a temperature of 1300 K and 3.2\%, 0.02 W/cm$^{2}$, when the emitter is assumed to have a temperature of 800 K.\par

\section{Summary}

We comment briefly on previous far-field TPV designs which have utilized Gallium Antimonide \cite{dashiell2006quaternary,lenert2014nanophotonic} as the photovoltaic cell for converting the thermal radiation into electric power \cite{bauer2011thermophotovoltaics}. Our analysis shows that the optimum cell design for near-field TPV is fundamentally different and the presence of van Hove Singularities in the material comprising the cell is critical for the spectrally selective nature of the transferred energy. This in turn should lead to fundamental improvements in energy conversion efficiency as shown by the Shockley Queisser analysis. The implementation of near-field TPV designs is more challenging than far-field TPV designs. However, a fundamental promise of near-field TPV, as mentioned before, is enhanced heat transfer for lower operating temperatures. This arises because the black body limit which fundamentally constraints far-field TPV does not apply to near-field TPV where heat exchange takes place due to tunneling of evanescent waves.

In summary, we have shown that the ideal joint density of electronic states for near-field photonic energy capture is mathematically equivalent to the Dirac delta distribution. This result immediately reveals why bulk semiconductors are clearly ill suited to near-field photonic energy capture. Thus switching to photovoltaic cells with van Hove singularities, seen in any semiconductor with a quantum-confined dimension, offers a clear path for improving the efficiency of future NFTPV devices. For experimental verification, we have provided model designs for two such low dimensional systems consisting of matched metamaterial CNT emitter absorber pairs and have shown that their Shockley-Queisser limit performance metrics are well beyond those achievable with current NFTPV device designs using bulk semiconductor photovoltaics.
Our main aim is to emphasize that the absorption characteristics of such low dimensional materials are ideal for near-field TPV and is an interesting avenue to explore in the development of thermophotovoltaic cells. Our work will also motivate future studies to adapt quantum well, quantum dot, two-dimensional semiconductors and CNT cells specifically for thermophotovoltaics.

\section{Acknowledgements}
This work has been supported by funding from the National Science and Engineering Research Council of Canada, Alberta Innovates Technology Futures, and the Helmholtz-Alberta initiative. We also thank Cris Cortes and Ward Newman for valuable discussion.
\section{Appendix: Outline for the calculation of the optical properties of CNTs}
As mentioned in the main text, the calculation of the optical properties of the (19,0) chiral vector CNTs used in our sample NFTPV designs has followed the Kubo formalism method described by Falkovsky and Varlamov\cite{Falkovsky2007}. The central equation for the dielectric permittivity tensor in this approach is
\begin{equation}
\epsilon_{ab}\left(k,\omega\right)=\epsilon_{sur}+\frac{e^2}{hc\epsilon_{o}}
\frac{P_{ab}\left(k,\omega\right)}{\omega},
\label{appendix_1}
\end{equation}
where
\begin{equation}
\begin{split}
 & P_{ab} \left(\omega,{\bf k}\right)=\int\frac{dp}{2\pi}\sum\limits_{\omega_{n}}\\
 & Tr\left[V_{a}\left(p\right)\mathcal{G}\left(\omega_n,p^+\right)V_b\left(p\right)\mathcal{G}\left(\omega_n,p^-\right)\right].	
\end{split}
\label{appendix_2}
\end{equation}
Here, $P_{ab}$ denotes the polarization matrix, with the subscripts $a$ and $b$ indicating directions, $\mathcal{G}$ the Matsubara Green function, $\omega$ the frequency of the exciting photon, $\omega_{n}$ the discrete frequencies of the Matsubara sum, $Tr$ the trace operation, $\epsilon_{o}$ the permittivity of free space, $\epsilon_{sur}$ the relative permittivity of the surrounding medium, $V_{a}$ the velocity operator matrix defined by the Heisenberg operator evolution equation 
\begin{equation}
V_{a}=\dot{x_{a}}=\frac{\left[x_{a},H\right]}{i\hbar}=\frac{\partial H}{\partial p_{a}},
\end{equation}$p$ the momentum integration variable, and the $+$ and $-$ superscripts the values $p+\frac{k}{2}$ and $p-\frac{k}{2}$ with $k$ standing for the momentum of the photon excitation. Moreover, the Green's function is explicitly defined as
\begin{equation}
\mathcal{G}=\left(i\omega_{n}-H\left(p\right)\right)^{-1},
\end{equation}
and its the sum over the discrete frequencies of the Matsubara sum, in terms of its matrix elements indexed by the subscripts $i$ and $j$, by
\begin{equation}
\sum\limits_{\omega_{n}}\mathcal{G}_{i}\left(\omega_n,p^+\right)\mathcal{G}_{j}\left(\omega_n,p^-\right)=\frac{f_o\left(\mathcal{E}_{j}\left(p^-\right)\right)-f_o\left(\mathcal{E}_i\left(p^+\right)\right)}
{\omega+i\eta-\left(\mathcal{E}_{j}\left(p^+\right)-\mathcal{E}_{i}\left(p^-\right)\right)}.
\end{equation}
In these definitions $H$ is the Hamiltonian of the system, $\eta$ the loss parameter following the relaxation time approximation, and $f_{o}\left(\mathcal{E}_{j}\right)$ the Fermi distribution at the energy level of the $j$ eigen state.\par
For the carbon nanotube system treated in the text, only $\pi$-orbital interactions have been considered due to the energy range of interest\cite{reich2002tight}. With these approximations the Hamiltonian is
\begin{equation}
H\left(p,j\right)=\gamma\left(e^{i\frac{i p a\sqrt{3}}{2}-\frac{i\pi j}{n}}+e^{i\frac{pa}{2\sqrt{3}}+\frac{i\pi j}{n}}+e^{-i\frac{pa}{\sqrt{3}}}\right),
\end{equation}
where $a$ is the lattice constant for graphene, $2.46\;\mathring{A}$, $j$ the angular number, $p$ the momentum of the electron along the tube, $n$ a constant related to the wrapping of the tube\cite{jorio2007carbon}, and $\gamma$ is the orbital overlap energy for graphene, $3.1\;eV$. The loss parameter $\eta=h/\tau$ is estimated with a relaxation of one picosecond\cite{Kohler2011,Kohler2013} for $\tau$. The energy of the nanotube is similarly defined as
\begin{equation}
\mathcal{E}_{j}=\pm\gamma\sqrt{1+4\;cos\left(\frac{\sqrt{3}p a}{2}\right)cos\left(\frac{j\pi}{n}\right)+4\;cos^{2}\left(\frac{j\pi}{n}\right)}.
\end{equation}

To extend the theory presented by Falkovsky and Varlamov from graphene to the single layer nanotubes, the Hamilitonian associated with graphene is first enlarged to take into account the energy level splitting resulting from the additional confinement in the nanotube. This step is accomplished by the substitution 
\begin{equation}
\left[ \begin{array}{ccc}
0 & H\\
H^* &0
\end{array} \right]
\rightarrow
\left[\begin{array}{cccc}
0& 0 & 0 & H_{\mathcal{E}1}\\
0& 0 & H_{\mathcal{E}2} & 0\\
0& H_{\mathcal{E}2}^* & 0& 0\\
H_{\mathcal{E}1}^* & 0& 0& 0
\end{array} \right],
\end{equation}
with $H$ as above, and the $\mathcal{E}_{1}$ and $\mathcal{E}_{2}$ subscripts denoting the lowest level energy bands for the particular chiral vector chosen following the discussion provided by Wong and Akinwade\cite{wong2010carbon}. The velocity operators for axial, $z$, and angular direction, $\theta$, are then:
\begin{equation}
V_{z}=\left[ \begin{array}{cccc}
0& 0 & 0 & v_{Z \mathcal{E}1}\\
0& 0 & v_{Z \mathcal{E}2} & 0\\
0& v_{Z \mathcal{E}2}^* & 0& 0\\
v_{Z \mathcal{E}1}^* & 0& 0& 0
\end{array} \right]
\end{equation}
for the axial direction, and
\begin{equation}
V_{\theta +}=\left[ \begin{array}{cccc}
0& 0 & v_{\theta \mathcal{E}2} & 0\\
0& 0 & 0& 0\\
v_{\theta \mathcal{E}2}^* &0 & 0& 0\\
0& 0& 0& 0
\end{array} \right],\\
V_{\theta -}=\left[ \begin{array}{cccc}
0& 0 & 0& 0\\
0& 0 & 0& v_{\theta \mathcal{E}1} \\
0&0 & 0& 0\\
0& v_{\theta \mathcal{E}1}^* & 0& 0
\end{array} \right]
\end{equation}
for increasing and decreasing angular momentum along the angular direction respectively, see\cite{jorio2007carbon}. Note that for the angular velocity operators the derivative is taken with respect to the natural discrete momenta. These operators are then converted to the eigenbasis of the Hamilionian by the linear transformation $v\rightarrow U^{\dagger}v U$, with
\begin{equation}
U=\left[ \begin{array}{cccc}
-\frac{H_{\mathcal{E}1}}{\left|H_{\mathcal{E}1}\right|}& 0 & 0&\frac{H_{\mathcal{E}1}}{\left|H_{\mathcal{E}1}\right|} \\
0&-\frac{H_{\mathcal{E}2}}{\left|H_{\mathcal{E}2}\right|} &\frac{H_{\mathcal{E}2}}{\left|H_{\mathcal{E}2}\right|}  & 0\\
0&1 & 1& 0\\
1& 0& 0& 1
\end{array} \right]. 
\end{equation}
An identical procedure is undertaken for the effective mass operators, $M_{ij}=\pm\frac{1}{\hbar^{2}}\frac{\partial^{2}H}{\partial p_{i}\partial p_{j}}$, for the calculation of the static conductivities used in the relaxation time approximation. With these substitutions the procedure detailed by Falkovsky and Varlamov\cite{Falkovsky2007}, equations \eqref{appendix_1} and \eqref{appendix_2}, can be followed directly to achieve Fig.4.  
\section{References}
\bibliography{QCNFTPV}
\end{document}